\documentclass{article}
\pdfoutput=1

\usepackage{subcaption}
\usepackage{siunitx}
\usepackage{physics}
\usepackage{graphicx,amsmath,amssymb,epsf} 
\usepackage{latexsym,bm,slashed} 
\usepackage{xcolor}
\usepackage{hyperref, cite} \hypersetup{colorlinks, citecolor=blue,
  filecolor=blue, linkcolor=magenta,
  urlcolor=purpure,hyperfootnotes=true,pdftex}

\title{The Dynamics of Particle-Particle Correlations and the Ridge Effect in Proton-Proton Collisions} 
\author{G. Cal\'e$^{\, 1,2}$, G. Chachamis$^{\, 2,3}$,  A. Sabio Vera$^{\, 4,5}$\\ 
\\
\small $^1$ Departamento de F\'isica, Universidade de Lisboa, Campo Grande
1749-016 Lisboa, Portugal.\\
\small $^2$ Laborat{\' o}rio de Instrumenta\c{c}{\~ a}o e F{\' \i}sica Experimental de Part{\' \i}culas (LIP),\\
\small Av. Prof. Gama Pinto, 2, P-1649-003 Lisboa, Portugal.\\
\small $^3$ Departamento de Estad\'istica, Inform\'atica y Matem\'aticas, \\
\small Universidad P\'ublica de Navarra-UPNA, 31006 Pamplona, Spain. \\
\small $^4$ Instituto de F{\'\i}sica Te{\'o}rica UAM/CSIC, Nicol{\'a}s Cabrera 15, E-28049 Madrid, Spain.\\
\small $^5$ Theoretical Physics Department, Universidad Aut{\' o}noma de Madrid, E-28049 Madrid, Spain.\\}

\begin{document}

\date{\today}

\maketitle

\begin{abstract}
%\setstretch{1.0}
In  high-energy particle physics, the study of particle-particle correlations in proton-proton and heavy-ion collisions constitutes a pivotal frontier in the effort to understand the fundamental dynamics of  the strong force.  To the best of our knowledge, we employ for the first time the BFKL dynamics implemented in a Monte Carlo code in momentum space to compute final state correlations in proton-proton collisions. Our present work aims to investigate whether the particular dynamics of the high-energy limit of QCD can contribute to the long-range rapidity correlations and the enigmatic ridge effect in proton-proton collisions.
\end{abstract}

\section{Introduction}
Particle-particle correlations constitute a central aspect of high-energy particle physics, as they enable the exploration of the intricate dynamics of strong force interactions in various collision systems.  It is a tool with a long history going back to 
1960 when Goldhaber, Goldhaber, Lee and Pais extracted from two-pion correlations the spatial extent of the annihilation fireball in proton-antiproton reactions~\cite{Goldhaber:1960sf} (for a review, see Ref.~\cite{Heinz:1999rw}).

Particle-particle correlations, in particular correlations in the pseudorapidity-azimuthal anlge plane, became anew an extremely important tool for high-energy particle physics in the last 20 years, initially in the analysis of the data from the Relativistic Heavy-Ion Collider (RHIC) at Brookhaven National Laboratory. There, while studying correlations of the final state particles in Au-Au collisions, it was observed the so-called ridge effect, where the produced particles appear as two "ridges" opposite in azimuthal angle $\phi$, with approximately flat rapidity distributions~\cite{STAR:2005ryu,Putschke:2007mi}.
The ridge in ion-ion collisions was seen as a collective phenomenon, namely as a signature of the formation of quark-gluon plasma (QGP) - a hot and dense state of matter~\cite{VanHove:1986mz,Ericson:138264,Heinz:2000bk}. The formation of the ridge was explained by the hydrodynamic expansion of the QGP, which creates a pressure gradient that drives the particles to flow along the direction of the collision axis. In other words, the long-range correlations in rapidity
for azimuthal angle differences near zero are usually attributed to a collective hydrodynamical flow due to an initial anisotropy 
in the collision of the two ions that survives in the distributions of the final-state particles through the collective expansion of the medium~\cite{Dusling:2015gta}.

Later, the ridge was also observed at the Large Hadron Collider (LHC) in smaller collision systems, such as proton-proton collisions. The first observation of the ridge effect in proton-proton collisions was reported by the CMS collaboration in 2010, using data from the LHC at a center-of-mass energy of 7 TeV~\cite{CMS:2010ifv}. Long-range correlations have been observed in high-multiplicity proton--proton (pp)~\cite{ATLAS:2015hzw,CMS:2015fgy}, proton--nucleus (pA)~\cite{ALICE:2012eyl,CMS:2012qk,ATLAS:2014qaj,CMS:2016est}, and light nucleus-nucleus collisions~\cite{PHENIX:2018lia,PHENIX:2017xrm}. These findings have raised the question on whether an explanation based on collective phenomena is adequate in hadronic collisions. The formation of a medium and its subsequent evolution, assumed to explain the ridge in heavy-ion collisions, might not apply in small collision systems, since the requirement of thermal equilibrium may not be fulfilled due to the small system size. 

Despite the huge theoretical and experimental effort, the ridge effect in proton-proton collisions is still not fully understood~\cite{Strickland:2018exs,Loizides:2016tew,Nagle:2018nvi} despite some recent progress~\cite{Schenke:2020mbo,Zhao:2022ugy,STAR:2022pfn}.
At present, most of the possible explanations in the literature  (for a very nice summary,  see Ref.~\cite{Altinoluk:2020wpf})  involve either the Color Glass Condensate (CGC) framework~\cite{Jalilian-Marian:1997jhx,Iancu:2002xk,Gelis:2010nm} and gluonic flux tubes~\cite{Dumitru:2008wn,Glazek:2018gqs} or hydrodynamic flow~\cite{Bzdak:2013zma,Werner:2010ss}, 
while numerous works offer various descriptions of possible mechanisms responsible for the ridge effect, see for example Refs.~\cite{Bjorken:2013boa,Li:2012hc,Armesto:2004pt,Majumder:2006wi,Romatschke:2006bb,Shuryak:2007fu,Dumitru:2010iy,Dusling:2012iga,Dusling:2012cg,Dusling:2013oia,Albacete:2013ei,Blok:2017pui,Blok:2018xes,Blok:2021aza,Musumeci:2023noi} and references therein.

Recently, a number of experimental studies have given valuable new information but have not resolved the mystery of the ridge effect in small systems. Collective behaviour of final-state hadrons was studied in high-multiplicity events at photoproduction and deep inelastic $e p$ scattering at a centre-of-mass energy of 318 GeV with the ZEUS detector at HERA~\cite{ZEUS:2021qzg}. In that study, neither the measurements in photoproduction processes nor those in neutral current deep inelastic scattering showed significant collective behaviour similar to what was observed in high-multiplicity hadronic collisions.
Furthermore, measurements of two-particle angular correlations of charged particles emitted in hadronic Z decays were presented in Ref.~\cite{Badea:2019vey}. The analysis was done with archived $e^+$ $e^{-}$ annihilation data at a center-of-mass energy of 91 GeV which were collected with the ALEPH detector at LEP.  There, no significant enhancement of long-range correlations was observed. A subsequent analysis on data with a center-of-mass energy of up to 209 GeV identified a long-range near-side excess in the correlation function when calculating particle kinematic variables with respect to the thrust axis~\cite{Chen:2023njr}. Very recently, ALICE released a study~\cite{ALICE:2023ulm} (see also Ref.~\cite{ALICE:2023lyr}) of the ridge yield measured in a hadronic system of similar multiplicity to the multiplicity of events that come from $e^+$ $e^{-}$ annihilation. The ridge yield was substantially larger than what was observed by the ALEPH analysis for center-of-mass energy of 91 GeV. Furthermore,  CMS has reported the observation of enhanced long-range elliptic anisotropies inside high-multiplicity jets in pp collisions~\cite{CMS:2023iam}. All these are quite indicative that the mechanisms for ridge yield production in very small hadronic systems are not understood and that more theoretical investigation is needed.

With the present work, we want to study whether the QCD dynamics that governs the hardest subprocesses in proton-proton collisions could be responsible for correlations at small azimuthal angles (near side) and large rapidity separations. Typically, in a fixed order calculation, the partonic cross-section is a very low multiplicity event before hadronization, two outgoing partons at leading order (LO), three outgoing partons at next-to-leading order (NLO) and four outgoing partons at next-to-next-to-leading order (NNLO). If the near side correlations were dictated by the hard scattering part, leaving aside the fact that probably one could easily theoretically compute them, we would be able to see them at any multiplicity events which is not the case. On the other hand, going beyond a fixed order calculation for the partonic cross-section, in particular in the high energy limit of QCD (more accurately, the multi-Regge limit), we know that we have the emergence of interesting effects regarding the dominant dynamics, such as the decoupling of the rapidity from the transverse degrees of freedom, the reggeization of the $t$-channel exchanged gluons and rapid increase of the amount of small-$x$ gluons with similar transverse momenta in the colliding protons. Actually, this rapid increase of the number of gluons in the protons eventually leads to unitarity violation at very high energies and the main mechanism to restore the latter is the introduction of parton saturation~\cite{Gribov:1983ivg}, a key concept within the framework of CGC.

In this paper, we use Monte Carlo simultations to compute the rapidity-azimuthal angle correlations for proton-proton collisions, in particular we use {\tt Pythia8}~\cite{Bierlich:2022pfr} as a base reference  and {\tt BFKLex}~\cite{Chachamis:2022jis,deLeon:2020myv,Chachamis:2015ico,Chachamis:2015zzp,Caporale:2013bva,Chachamis:2012qw,Chachamis:2012fk,Chachamis:2011nz,Chachamis:2011rw} which is a Monte Carlo code that generates the hard scattering part of the collision using the Balitsky-Fadin-Kuraev-Lipatov (BFKL) resummation framework~\cite{Lipatov:1985uk,Balitsky:1978ic,Kuraev:1977fs,Kuraev:1976ge,Lipatov:1976zz,Fadin:1975cb}. In particular, it employs the iterative solution of the BFKL equation cast in a suitable form for Monte Carlo studies directly in momentum space~\cite{Schmidt:1996fg}. Both computations are performed within the collinear factorization scheme~\cite{Collins:1985ue,Collins:1989gx} where the partonic cross section is convoluted with the parton distribution functions (PDF) of the proton. What we aim at with this study is to find out whether any near side correlations arise once we switch from a [LO matrix elements] $+$ [parton shower] approach to a BFKL-based calculation at leading logarithmic accuracy. The importance of computing the correlation distribution from both approaches directly in momentum space from final-state jets and minijets cannot be overstated, as it allows us to avoid any modelling considerations.

In the next section, we give a short introduction on BFKL and {\tt BFKLex}. In Section 3, we lay the groundwork for our study, we present our results and discuss our findings. In Conclusion, we offer our final remarks and provide insights for future research directions.

\section{BFKL and {\tt BFKLex}}

An important line of research within particle phenomenology at colliders 
is to search for effects that are associated with the high energy limit
of QCD and to pin down observables that can reveal the effects of the BFKL domain~\cite{Lipatov:1985uk,Balitsky:1978ic,Kuraev:1977fs,Kuraev:1976ge,Lipatov:1976zz,Fadin:1975cb}. 
This has proven to be a rather challenging task since the typical phenomenological calculations based on 
matrix elements computed at fixed order along with the Dokshitzer-Gribov-Lipatov-Altarelli-Parisi (DGLAP) 
evolution~\cite{Gribov:1972ri,Gribov:1972rt,Lipatov:1974qm,Altarelli:1977zs,Dokshitzer:1977sg} 
to account for the PDFs tend to describe the bulk of the data adequately.

The key idea in this formalism is that, when the center-of-mass energy $\sqrt{s} \to \infty$,
Feynman diagrams that contribute terms of the form {$\alpha_s^n 
\log^n{\left(s\right)} \sim \alpha_s^n \left(y_A-y_B\right)^n$} give the dominant numerical
contributions to the computation of cross-sections.  $y_{A}$ and $y_{B}$ are the rapidities of  some properly chosen
tagged particles or  jets in the final state,  such
that their rapidity difference $Y = y_A - y_B$ is the largest among the particles or jets in the final state. 
The terms  {$\alpha_s^n 
\log^n{\left(s\right)} $ can be of order unity and 
therefore, these diagrams must be resummed in order to accurately describe experimental observables. In this limit,
a decoupling between transverse and longitudinal degrees of freedom takes place which allows to evaluate cross sections in the factorized form: 
\begin{eqnarray}
\sigma^{\rm LL} &=& \sum_{n=0}^\infty {\cal C}_n^{\rm LL}  \alpha_s^n 
\int_{y_B}^{y_A} d y_1 \int_{y_B}^{y_1} d y_2 \dots \int_{y_B}^{y_{n-1}} d y_n \nonumber\\ 
&=& \sum_{n=0}^\infty \frac{{\cal C}_n^{\rm LL}}{n!} 
\underbrace{\alpha_s^n \left(y_A-y_B\right)^n }_{\rm LL} \nonumber
\end{eqnarray}
where LL stands for the  leading log approximation and $y_i$ correspond to the rapidity of emitted particles. The LL BFKL formalism allows one to calculate the coefficients ${\cal C}_n^{\rm LL}$~\cite{Lipatov:1985uk,Balitsky:1978ic,Kuraev:1977fs,Kuraev:1976ge,Lipatov:1976zz,Fadin:1975cb}. 
The next-to-leading log approximation (NLL)~\cite{Fadin:1998py,Ciafaloni:1998gs} is 
much more complicated since it is sensitive to the running of the strong coupling and  to the choice of energy scale in the logarithms.
One can parametrize the freedom in the choice of these two scales, respectively, by introducing the constants 
${\cal A}$ and ${\cal B}$ in the previous expression: 
\begin{eqnarray}
\sigma^{LL+NLL} &=& 
\sum_{n=1}^\infty \frac{{\cal C}_n^{\rm LL} }{n!}  \left(\alpha_s- {\cal A} \alpha_s^2\right)^n \left(y_A-y_B - {\cal B}\right)^n \nonumber\\
&&\hspace{-2.4cm}= \sigma^{\rm LL}  - \sum_{n=1}^\infty   \frac{\left({\cal B}  \, {\cal C}_n^{\rm LL} +  (n-1) \, {\cal A} 
\, {\cal C}_{n-1}^{\rm LL} \right)}{(n-1)!}  \underbrace{ \alpha_s^n 
\left(y_A-y_B\right)^{n-1}}_{\rm NLL} + \dots \nonumber
\end{eqnarray}
We see that at NLL a power in $\log{s}$ is lost w.r.t.\ the power of the coupling. Within the formalism, we can then calculate cross sections using the following factorization formula (with $Y\simeq \ln{s}$)
 \begin{eqnarray}
\sigma (Q_1,Q_2,Y) = \int d^2 \vec{k}_A d^2 \vec{k}_B \, \underbrace{\phi_A(Q_1,\vec{k}_A) \, 
\phi_B(Q_2,\vec{k}_B)}_{\rm PROCESS-DEPENDENT} \, \underbrace{f (\vec{k}_A,\vec{k}_B,Y)}_{\rm UNIVERSAL}, \nonumber
\end{eqnarray}
where $\phi_{A,B}$ are process-dependent impact factors which are functions of some external scale, $Q_{1,2}$, and some internal momentum for reggeized gluons, $\vec{k}_{A,B}$.  The gluon Green's function $f$ is universal, it depends on $\vec{k}_{A,B}$ and on the colliding energy of the process $\sim e^{Y/2}$. 
It corresponds to the solution of the BFKL equation. In momentum space, the BFKL equation  at LL reads
\begin{equation}
\omega \, f_\omega\left(\vec{k}_A, \vec{k}_B\right)=\delta^2\left(\vec{k}_A-\vec{k}_B\right)+\int \mathrm{d}^2 \vec{k}\,\,\, \mathcal{K}\left(\vec{k}_A, \vec{k}\right)\,\, f_\omega\left(\vec{k}, \vec{k}_B\right)\,,
\end{equation}
where $\mathcal{K}\left(\vec{k}_a, \vec{k}\right)$ is the BFKL kernel
\begin{equation}
\mathcal{K}\left(\vec{k}_a, \vec{k}\right)=\underbrace{2 \omega\left(-\vec{q}^2\right) \delta^2\left(\vec{k}_a-\vec{k}\right)}_{\mathcal{K}_{\text {virt }}}+\underbrace{\frac{N_c \alpha_s}{\pi^2} \frac{1}{\left(\vec{k}_a-\vec{k}_b\right)^2}}_{\mathcal{K}_{\text {real }}} .
\end{equation}
The solution of the BFKL equation at LL in transverse momentum representation can be written
in an iterative form~\cite{Schmidt:1996fg} as
\begin{eqnarray}
f &=& e^{\omega \left(\vec{k}_A\right) Y}  \Bigg\{\delta^{(2)} \left(\vec{k}_A-\vec{k}_B\right) + \sum_{n=1}^\infty \prod_{i=1}^n \frac{\alpha_s N_c}{\pi}  \int d^2 \vec{k}_i  
\frac{\theta\left(k_i^2-\lambda^2\right)}{\pi k_i^2} \nonumber\\
&&\hspace{-1.2cm}\int_0^{y_{i-1}} \hspace{-.3cm}d y_i e^{\left(\omega \left(\vec{k}_A+\sum_{l=1}^i \vec{k}_l\right) -\omega \left(\vec{k}_A+\sum_{l=1}^{i-1} \vec{k}_l\right)\right) y_i} \delta^{(2)} 
\left(\vec{k}_A+ \sum_{l=1}^n \vec{k}_l - \vec{k}_B\right)\Bigg\}, \nonumber
 \end{eqnarray}
where the gluon Regge trajectory reads
\begin{eqnarray}
\omega \left(\vec{q}\right) &=& - \frac{\alpha_s N_c}{\pi} \log{\frac{q^2}{\lambda^2}} \nonumber
\end{eqnarray}
and $\lambda$ is a regulator of infrared divergencies. This solution has been studied at length in a series of papers and it served as the basis in order to construct the Monte Carlo event code {\tt BFKLex} which has had multiple 
applications in collider phenomenology and more formal studies~\cite{Chachamis:2022jis,deLeon:2020myv,Chachamis:2015ico,Chachamis:2015zzp,Caporale:2013bva,Chachamis:2012qw,Chachamis:2012fk,Chachamis:2011nz,Chachamis:2011rw}. In this paper, we will run {\tt BFKLex} to LL acuracy.

\section{Results and Discussion}
Two-particle correlations are analyzed in a two-dimensional azimuthal $\Delta \eta$-$\Delta \phi$ phase space, where $\Delta \eta$ and $\Delta \phi$ denote the difference of  the pseudorapidity $\eta$ and the azimuthal angle $\phi$  of the two selected particles, respectively. The two-particle correlation function is defined as
\begin{equation}
C(\Delta \eta, \Delta \phi)=\frac{S(\Delta \eta, \Delta \phi)}{B(\Delta \eta, \Delta \phi)}\,,
\end{equation}
where $S$ and $B$ denote particle pair distributions from the same event and from different events respectively, 
representing the signal and background contributions, see for example Ref.~\cite{PHENIX:2008osq}.
As we mentioned in the introduction, one doesn't expect to notice any significant type of near-side correlations just by studying the outgoing partons in a fixed order computation setup. More specifically, since at LO, we have only two outgoing partons flying back to back, we expect these to contribute to the far side ridge since their azimuthal angle difference will be around $\pi$. For {\tt Pythia8}, we need to switch on the initial (ISR) and final state radiation (FSR) allowing parton shower. We also switch on multiple partonic interactions (MPI). This will largely increase the number of the quarks and gluons that will eventually hadronise and give the final state hadrons that are typically detected by the LHC experiments. {\tt BFKLex} does not have parton shower implemented.  However, we know from the iterative solution described in Section 2 that for medium and large rapidity separations $\Delta \eta = \eta_a - \eta_b$ between the most forward ($\eta_a$) and most backward ($\eta_b$) partons in BFKL evolution, we have a significant amount of emitted gluons (typically of the order of 15-20) with rapidities $\eta_i$: $\eta_b  < \eta_i < \eta_a$. Furthermore, we will not switch on any MPI scenario when we run {\tt BFKLex}.

In order to be able to qualitatively compare the correlations we compute from {\tt Pythia8}  and {\tt BFKLex}, we need to employ the following method: we switch on parton shower in {\tt Pythia8} and we use the anti-k$_T$ clustering algorithm~\cite{Cacciari:2008gp} as implemented in {\tt fastjet}~\cite{Cacciari:2011ma,Cacciari:2005hq} to cluster the final state partons (just before hadronization) into minijets. The term minijet was initially used to describe emissions of gluons in the BFKL framework. More precisely, minijets was the term used in the study of dijet production --when the final state is characterized by two jets widely separated in rapidity-- to describe any emission activity between the two bounding jets. In particular, we define a minijet as a jet entity with a rapidity value anywhere between the most forward and backward jet rapidities and transverse momentum that can either be large enough e.g.~$p_T^{\text{minijet}}> 20$ GeV, to enable experimental detection (thus qualifying as a jet) or much smaller, approaching a few GeV, in which case it appropriately earns its designation as a minijet.

There are two scales we need to set up for the runs. The first one is $\hat{p}_{T,\text{min}}$, which is the lowest allowed $p_T$ for the outgoing
partons in the partonic scattering, and the second is $p^{\text{jet}}_{T,\text{min}}$ , which defines the minimum allowed $p_T$ for a minijet. 
The former is a parameter passed to the generation of events in {\tt Pythia} whereas the latter is passed to  {\tt fastjet} for the clustering.  For the hard scattering part of the collision in {\tt BFKLex}, we impose a similar to $\hat{p}_{T,\text{min}}$ low cutoff, which controls the lowest allowed $p_T$ of
the most forward and most backward jet or minijet. We decided to use the default PDF set of {\tt Pythia} for all runs in the paper, namely, 
NNPDF2.3 QCD$+$QED LO~\cite{Ball:2013hta}. Despite the fact that is not possible to conduct a realistic experimental analysis on minijets, their study remains of great importance at the theoretical level, offering invaluable insights into fundamental processes. 

For the runs with {\tt BFKLex}, we allow the emission of up to 20 gluons in order to compute the gluon Green's function that governs the partonic cross section in the BFKL setup. One could allow for events with an even larger number of final state gluons but such events contribute very little to the total cross section and they do not change the correlation in any visible way. Next, we use again the anti-k$_T$ algorithm to cluster these gluons into minijets. That way, the rapidity-azimuthal angle correlations calculated for minijets above some lower $p_T$ cutoff are infrared safe observables and we can compare the respective distributions from the two Monte Carlo codes.

At this point, it is important to ascertain that the qualitative attributes of the correlation distributions remain intact during the transition from final state minijets to final state partons in our analysis. While it is generally expected that the correlation distributions exhibit similar characteristics when computed for both minijets and final state partons  (after parton shower), it is crucial to empirically verify this. To this end, we employ {\tt Pythia8} as a benchmark reference, we calculate the correlations for minijets and for partons and we verify that indeed we observe similar distributions with a noteworthy difference  in the region near  $\Delta \phi \sim 0\,\,, \Delta \eta \sim 0$,
that can be explained easily due the clustering effects as we will show in the following. We should also note here that final state partons will undergo hadronization, introducing additional variables to the overall dynamics. Nonetheless, given our current focus on investigating whether the near side ridge originates primarily from dominant QCD dynamics prior to hadronization, concerns regarding hadronization effects are deemed secondary.

In Fig.~\ref{fig:pythia}, we present the correlation distribution computed for the final state partons (Left) and the minijets (Right). We observe that there are no long-range correlations in rapidity near $\Delta \phi \sim 0$. This can be seen more clearly in Fig.~\ref{fig:pythiaAngl} where we integrate over $\Delta \eta$ excluding the range $-2 < \Delta \eta < 2$ to avoid the peak  near $\Delta \phi \sim 0\,\,, \Delta \eta \sim 0$  due to correlations between partons that belong to the same minijet.  To be precise, the peak is a characteristic of the parton correlation distribution which turns into a dip once we compute the minijet correlation distribution since there cannot be two minijets closer than $R = 0.5$ in rapidity and azimuthal angle since in that case they would be clustered
by {\tt fastjet} into one minijet.

\begin{figure}[!h]
    \begin{center}
    \includegraphics[width=0.49\textwidth]
    {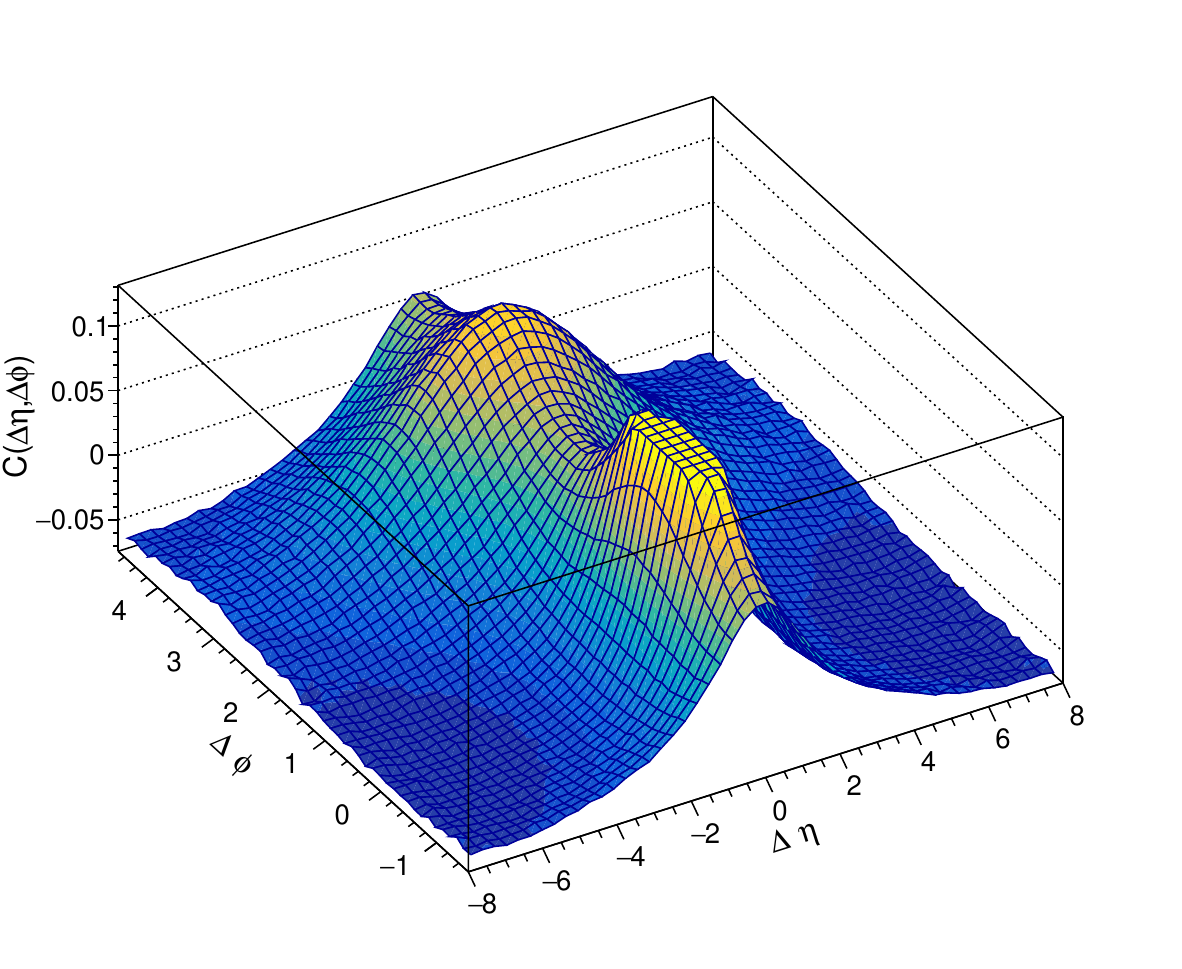}
    \includegraphics[width=0.49\textwidth]
    {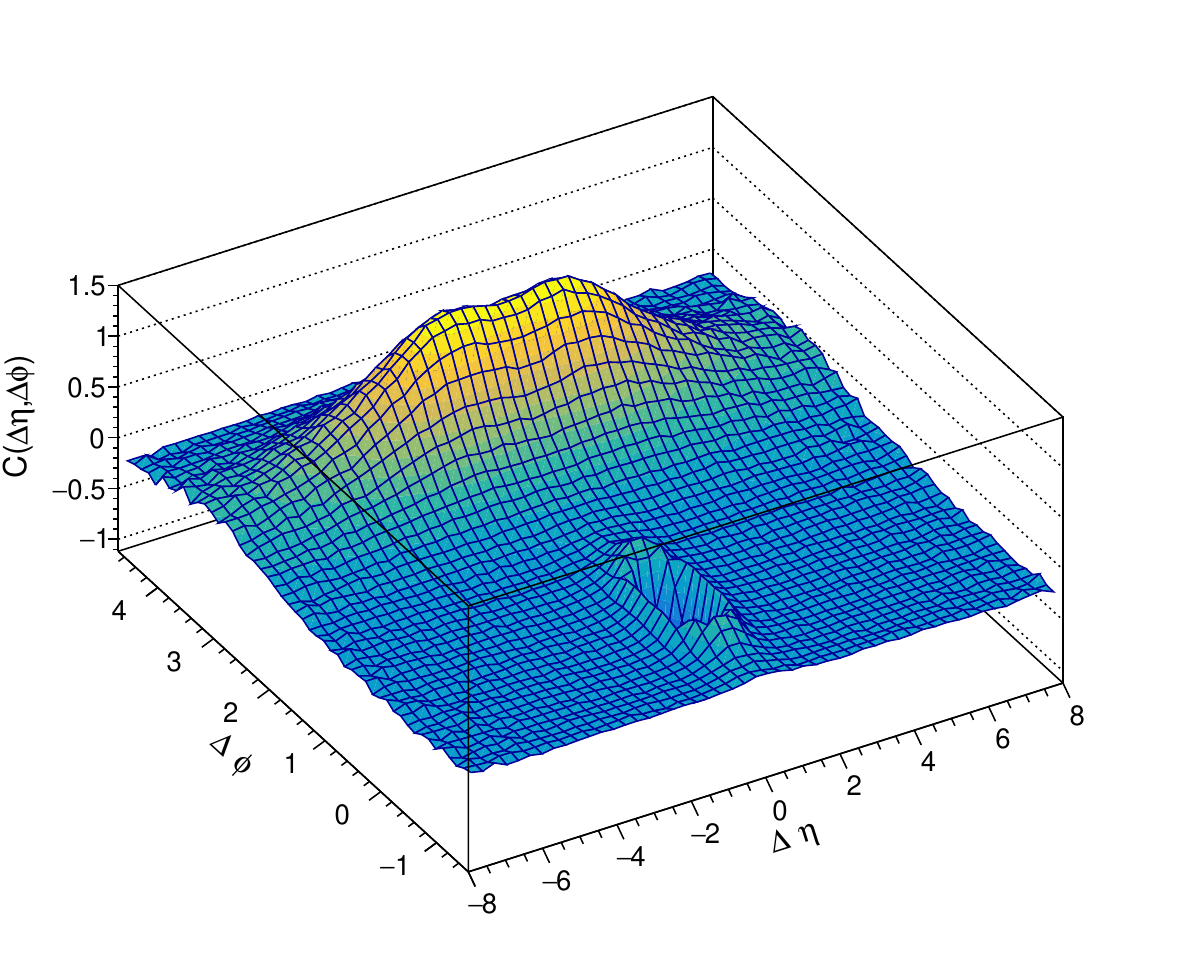}    
    \caption{Left: Parton correlation distribution obtained from p-p collisions with {\tt Pythia8}, with a minimum hard process transverse momentum cutoff of $ \hat{p}_{T,\text{min}}= 5$ GeV. Right: Minijet correlation distribution obtained from p-p collisions at {\tt Pythia8}, with a minimum hard process transverse momentum cutoff of $\hat{p}_{T,\text{min}}= 5$ GeV. The low $p_T$ cutoff in {\tt fastjet} was set to $p^{\text{jet}}_{T,\text{min}}= 5$ GeV and the jet radius to $R = 0.5$.}
    \label{fig:pythia}
    \end{center}
\end{figure}

\begin{figure}[!h]
    \begin{center}
    \includegraphics[width=0.49\textwidth]
    {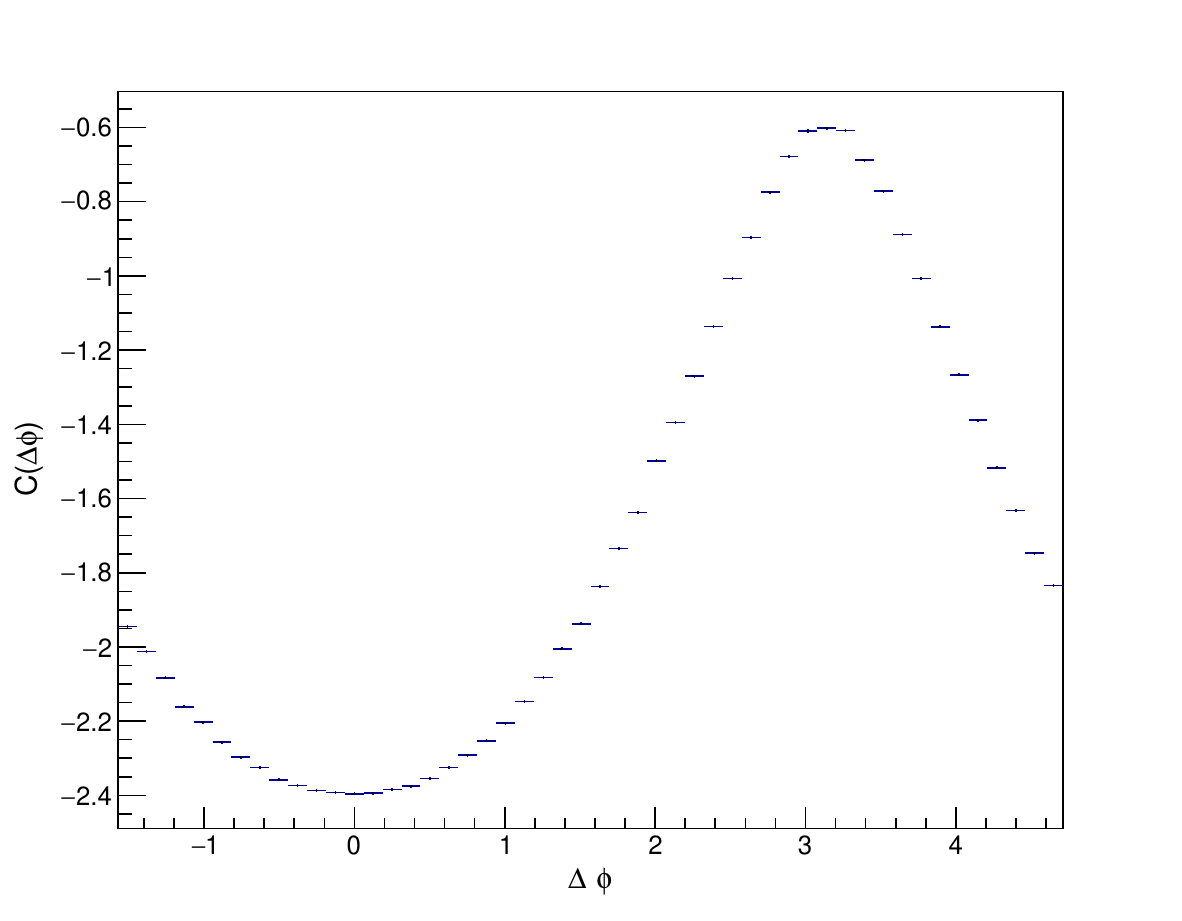}
    \includegraphics[width=0.49\textwidth]
    {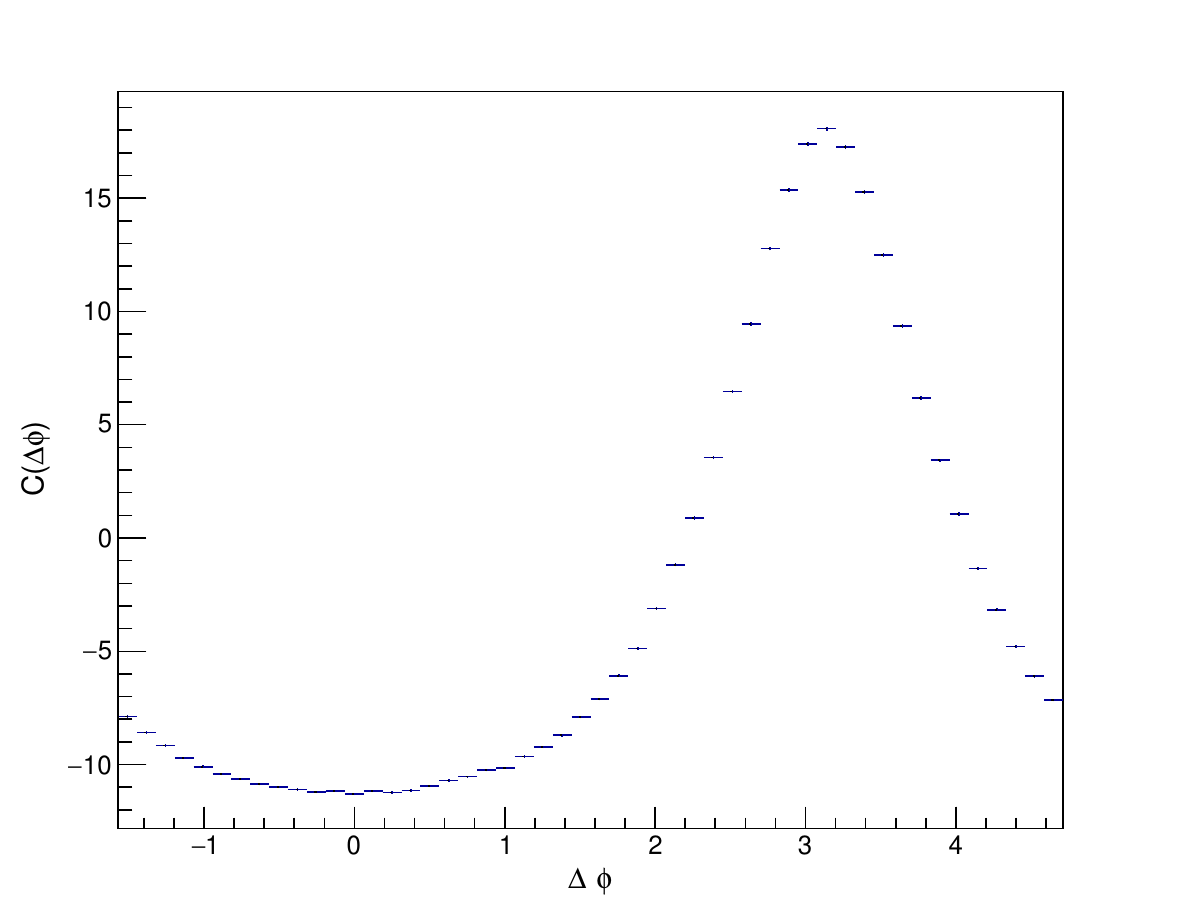}    
    \caption{Left: Parton azimuthal correlation distribution obtained after integrating the distribution in Fig.~\ref{fig:pythia} (Left) over $|\Delta \eta| > 2$. Right: Minijet azimuthal correlation distribution obtained after integrating the distribution in Fig.~\ref{fig:pythia} (Right) over $|\Delta \eta| > 2$.}
    \label{fig:pythiaAngl}
    \end{center}
\end{figure}

In Fig.\ref{fig:bfkl}, we present the correlation distribution computed for the minijets with {\tt BFKLex}. We impose a lower $p_T$ cutoff to the most forward and most backward jets, $ p_{T,\text{min}}= 5$ GeV. We use two different values for the low $p_T$ cutoff in {\tt fastjet}, $p^{\text{jet}}_{T,\text{min}}= 5$ GeV (Fig.~\ref{fig:bfkl}, Left) and $p^{\text{jet}}_{T,\text{min}}= 10$ GeV (Fig.~\ref{fig:bfkl}, Right).
As in the case with {\tt Pythia8}, we observe that there are no long-range correlations in rapidity near $\Delta \phi \sim 0$. In Fig.~\ref{fig:bfklAngl}, we integrate over $\Delta \eta$ excluding the range $-2 < \Delta \eta < 2$ to avoid the dip  near $\Delta \phi \sim 0\,\,, \Delta \eta \sim 0$. We also exclude the intervals $-8 < \Delta \eta < -7$ and  $7 < \Delta \eta < 8$ to avoid the statistical fluctuations in Fig.~\ref{fig:bfkl} that would only introduce unnecessary noise. We see that the distributions in Fig.~\ref{fig:bfklAngl} are very similar to the ones shown in Fig.~\ref{fig:pythiaAngl} (mainly Right).

\begin{figure}[!h]
    \begin{center}
    \includegraphics[width=0.49\textwidth]
    {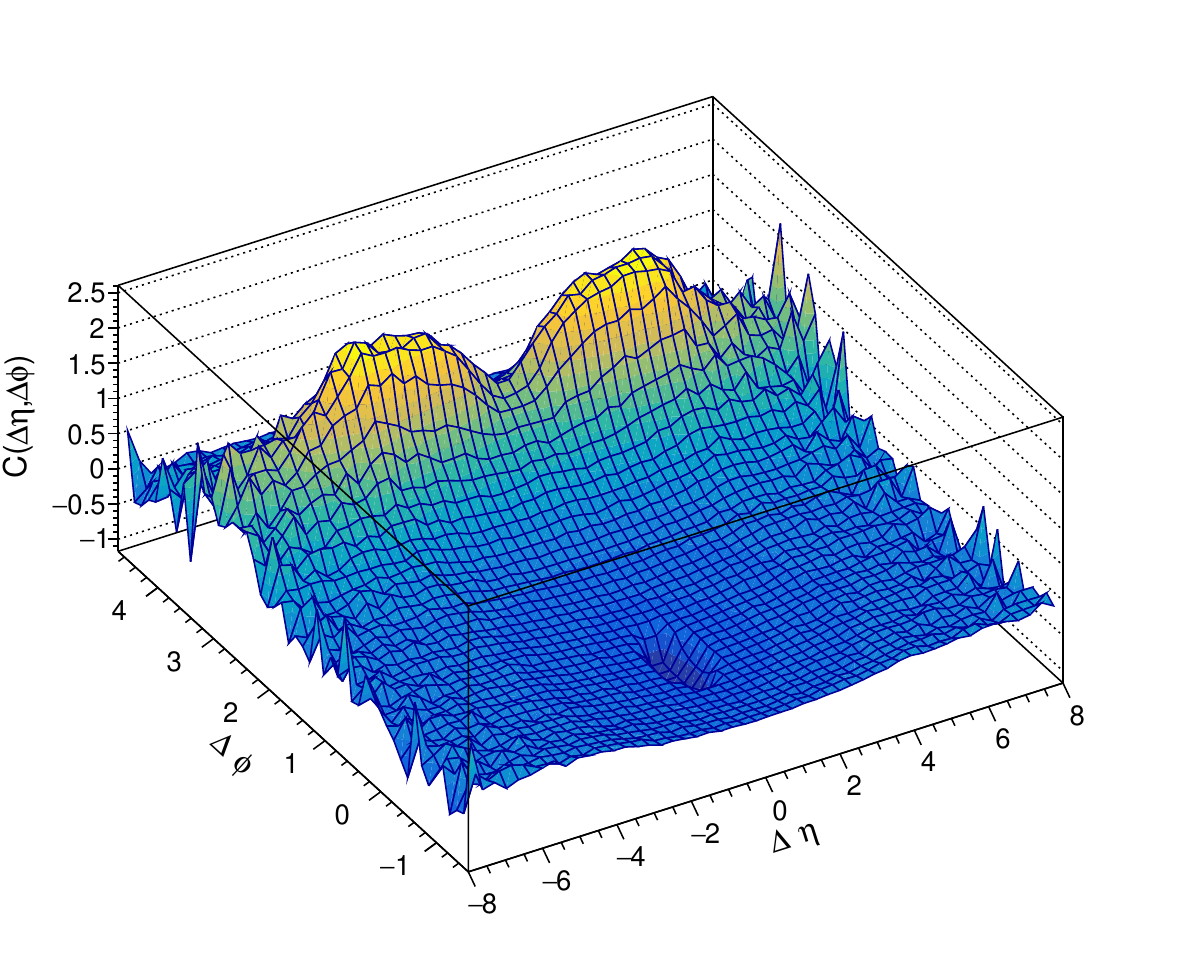}
    \includegraphics[width=0.49\textwidth]
    {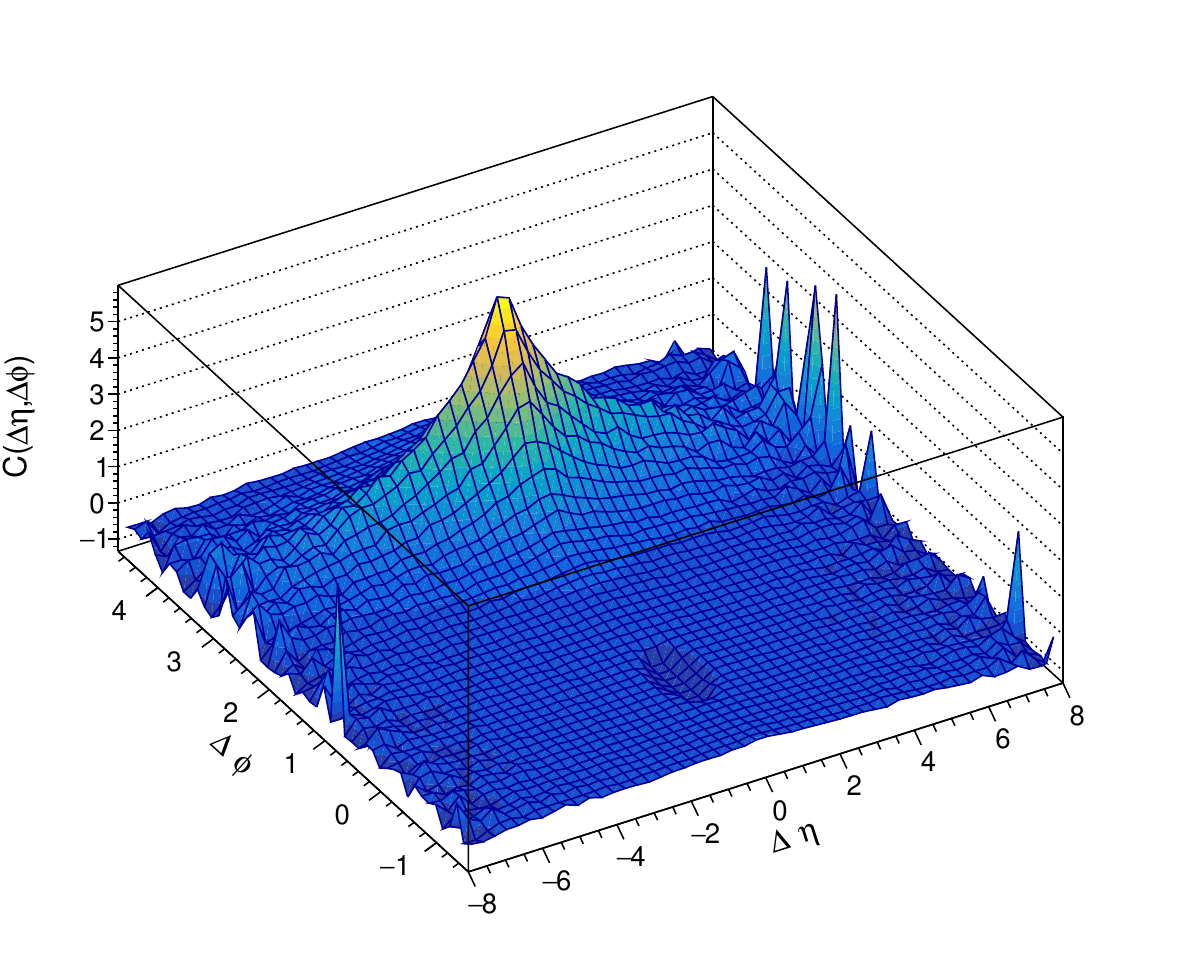}    
    \caption{Left: Minijet angular correlation distribution obtained from p-p collisions with {\tt BFKLex}. The minimum transverse momentum cutoff of the outermost in rapidity jets was set to $ p_{T,\text{min}}= 5$ GeV. The low $p_T$ cutoff in {\tt fastjet} was set to $p^{\text{jet}}_{T,\text{min}}= 5$ GeV and the jet radius to $R = 0.5$. Right: The same as to the left with the only difference that the  low $p_T$ cutoff in {\tt fastjet} was set to $p^{\text{jet}}_{T,\text{min}}= 10$ GeV.}
    \label{fig:bfkl}
    \end{center}
\end{figure}

\begin{figure}[!h]
    \begin{center}
    \includegraphics[width=0.49\textwidth]
    {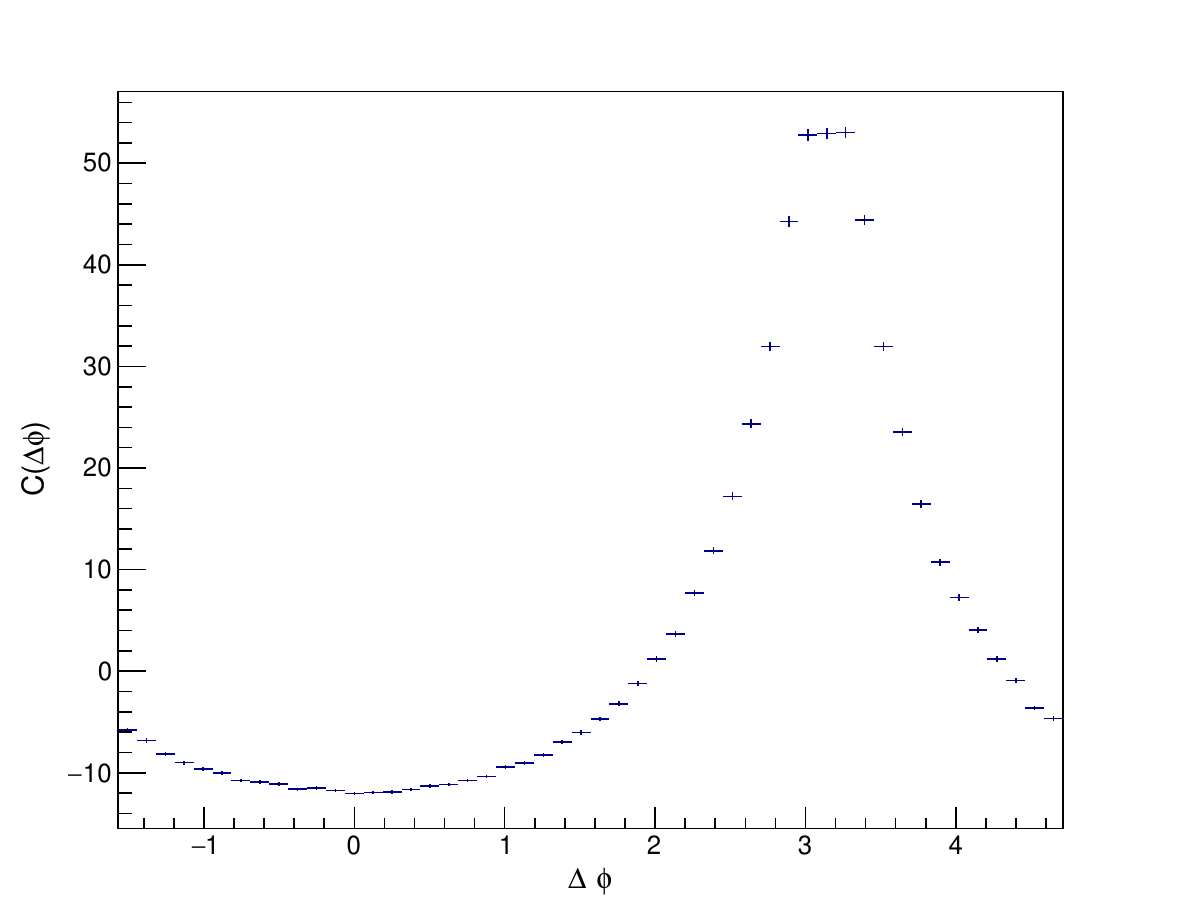}
    \includegraphics[width=0.49\textwidth]
    {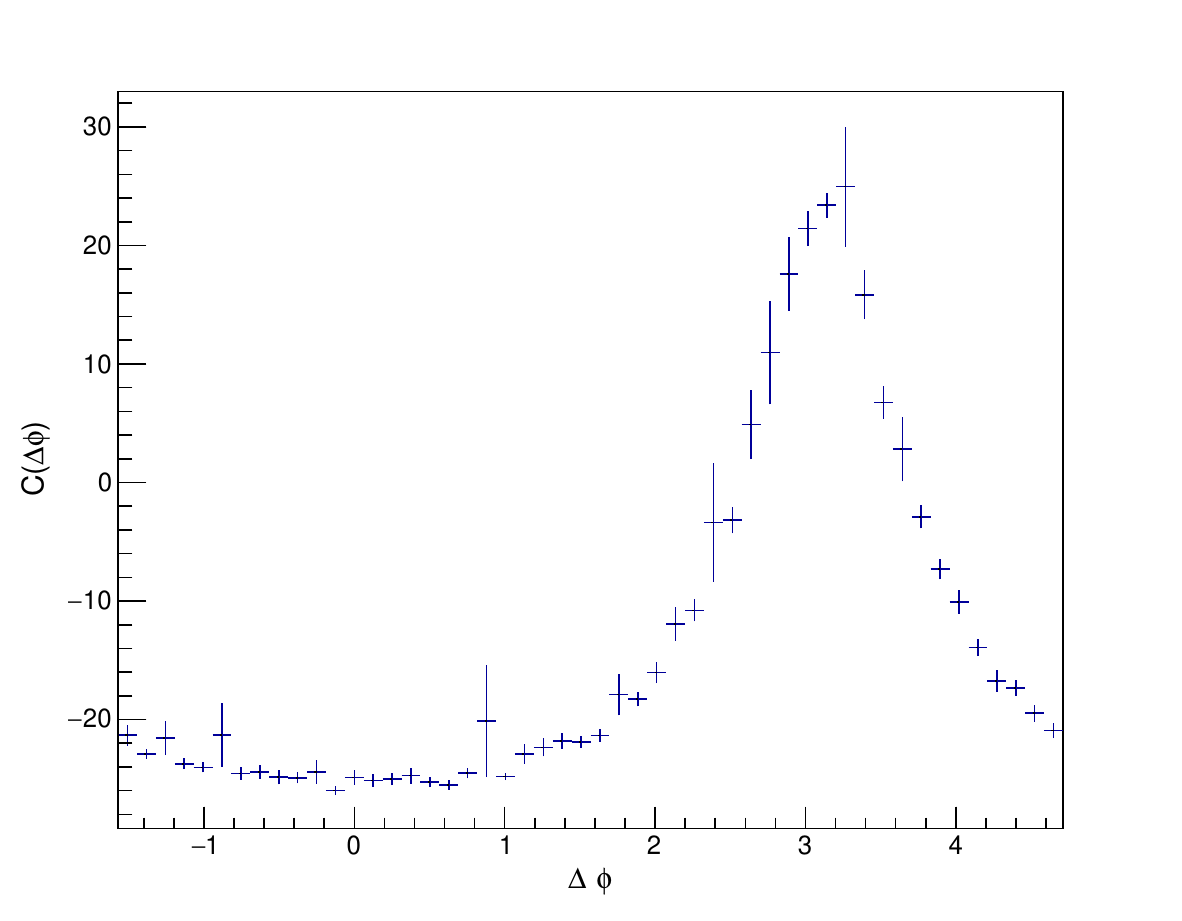}    
    \caption{Left: Minijet azimuthal correlation distribution obtained after integrating the distribution in Fig.~\ref{fig:bfkl} (Left) over $2 < |\Delta \eta| < 7$. Right: Minijet azimuthal correlation distribution obtained after integrating the distribution in Fig.~\ref{fig:bfkl} (Right) over $2 < |\Delta \eta| < 7$.}
    \label{fig:bfklAngl}
    \end{center}
\end{figure}

\section{Conclusion}

We presented the first BFKL based Monte Carlo study, directly in momentum space, of the rapidity-azimuthal angle correlations between minijets in proton-proton collisions, together with a similar study with {\tt Pythia}, the latter serving as a base reference.  
We found no indication that the specific dynamics of the high-energy limit of QCD are responsible for the long-range rapidity correlations and the puzzling ridge effect in small systems. It remains to be seen whether this conclusion will change with the inclusion of higher-order corrections in a full NLL accuracy BFKL analysis or with the incorporation of non-linear effects such as parton saturation.

\section*{Acknowledgements}

We thank Ma\l{}gorzata Anna Janik for useful discussions. This work has been supported by the Spanish Research Agency (Agencia Estatal de Investigaci{\'o}n) through the grant IFT Centro de Excelencia Severo Ochoa SEV-2016-0597, and by the Spanish Government grant FPA2016-78022-P. It has also received funding from the European Union's Horizon 2020 research and innovation programme under grant agreement No.\ 824093. The work of G.Cal\'e and G.Chachamis was supported by the Funda\c{c}{\~ a}o para a Ci{\^ e}ncia e a Tecnologia (Portugal) under project EXPL/FIS-PAR/1195/2021 (http://doi.org/10.54499/EXPL/FIS-PAR/1195/2021). G. Chachamis was also supported by the Funda\c{c}{\~ a}o para a Ci{\^ e}ncia e a Tecnologia (Portugal) under project CERN/FIS-PAR/0032/2021 (http://doi.org/10.54499/CERN/FIS-PAR/0032/2021) and contract `Investigador FCT - Individual Call/03216/2017'.

%\section*{References}
%\bibliographystyle{plainnat}
\bibliographystyle{ieeetr}
\bibliography{ridge-1.1} % Replace with the actual name of your .bib file

\end{document}